\PassOptionsToPackage{x11names,dvipsnames,table}{xcolor}
\documentclass[10pt, letterpaper, conference, compsocconf]{IEEEtran}
\def\BibTeX{{\rm B\kern-.05em{\sc i\kern-.025em b}\kern-.08em
    T\kern-.1667em\lower.7ex\hbox{E}\kern-.125emX}}

\usepackage{graphicx}

\usepackage[
    disable,     
    textwidth=.9\marginparwidth,
    colorinlistoftodos,
    prependcaption,
    linecolor=green,backgroundcolor=green!25,bordercolor=green
]{todonotes}

\makeatletter
\if@todonotes@disabled
\else 
\paperwidth=\dimexpr \paperwidth + 3.65cm\relax
\oddsidemargin=\dimexpr\oddsidemargin + 2cm\relax
\evensidemargin=\dimexpr\evensidemargin + 2cm\relax
\marginparwidth=\dimexpr \marginparwidth + 2.5cm\relax
\let\oldtitle\title
\renewcommand{\title}[1]{\oldtitle{#1\(_{\colorbox{red!20}{\tiny disable todos}}\)}}
\fi
\makeatother

\usepackage{mathtools}
\usepackage{adjustbox}
\usepackage{url}
\usepackage[capitalise]{cleveref}

\begin{document}

\title{FunLess: Functions-as-a-Service for\\ Private Edge Cloud Systems
}

\author{
\IEEEauthorblockN{Giuseppe De Palma\IEEEauthorrefmark{1}\IEEEauthorrefmark{2},
Saverio Giallorenzo\IEEEauthorrefmark{1}\IEEEauthorrefmark{2},
Jacopo Mauro\IEEEauthorrefmark{3},
Matteo Trentin\IEEEauthorrefmark{1}\IEEEauthorrefmark{2}\IEEEauthorrefmark{3}
Gianluigi Zavattaro\IEEEauthorrefmark{1}\IEEEauthorrefmark{2}
}
\IEEEauthorblockA{\IEEEauthorrefmark{1}
  \textit{Alma Mater Studiorum - Universit\`a di Bologna, Italy}
}
\IEEEauthorblockA{\IEEEauthorrefmark{2}
  \textit{INRIA, Sophia-Antipolis, France}
}
\IEEEauthorblockA{\IEEEauthorrefmark{3}
  \textit{University of Southern Denmark, Denmark}
}
}

\maketitle

\begin{abstract}
We present FunLess, a Function-as-a-Service (FaaS) platform tailored for the
private edge cloud system. FunLess responds to recent trends that advocate for
extending the coverage of serverless computing to private edge cloud systems and
enhancing latency, security, and privacy while improving resource usage. Unlike
existing solutions that rely on containers for function invocation, FunLess
leverages WebAssembly (Wasm) as its runtime environment. Wasm's lightweight,
sandboxed runtime is crucial to have functions run on constrained devices at the
edge. Moreover, the advantages of using Wasm in FunLess include a consistent
development and deployment environment for users and function portability (write
once, run everywhere)

We validate FunLess under different deployment scenarios, characterised by the
presence/absence of constrained-resource devices (Raspberry Pi 3B+) and the
(in)accessibility of container orchestration technologies---Kubernetes. We
compare FunLess with three production-ready, widely adopted open-source FaaS
platforms---OpenFaaS, Fission, and Knative.
Our benchmarks confirm that FunLess is a proper solution for FaaS private edge
cloud systems since it achieves performance comparable to the considered FaaS
alternatives while it is the only fully-deployable alternative on
constrained-resource devices---thanks to its small memory footprint.

\end{abstract}


\section{Introduction}
\label{intro}
The advent of serverless computing~\cite{BS19} introduced a paradigmatic shift
in the development of distributed systems, called Function-as-a-Service (FaaS).
In FaaS, programmers write and compose stateless functions, leaving to the
serverless platform the management of deployment and scaling. FaaS was first
proposed as a deployment modality for cloud architectures~\cite{BS19} that
pushed to the extreme the per-usage model of cloud computing, letting users pay
only for the computing resources used at each function invocation.

\noindent\emph{Private Edge Cloud FaaS}.\hspace{.5em}
While public clouds are the birthplace of serverless computing, recent
industrial and academic proposals demonstrated the desirability, benefits and
feasibility of moving FaaS outside public clouds. These solutions are tailored
for private and hybrid cloud scenarios~\cite{CIMSe23} and include
edge~\cite{baresi2019towards} and Internet-of-Things~\cite{SKM22} components.
From the industrial point of view, several FaaS platforms are designed for edge
computing (e.g., AWS
Greengrass\footnote{\url{https://aws.amazon.com/greengrass/}.}, Cloudfare
Workers\footnote{\url{https://www.cloudflare.com/en-gb/learning/serverless/glossary/serverless-and-cloudflare-workers/}.}).

In contrast to public edge-cloud computing solutions, \emph{private edge cloud
      systems} have the benefit of further reducing latency, increasing security and
privacy, and improving bandwidth and usage of high-end devices~\cite{SKM22}.
More precisely, private edge cloud systems are small-scale cloud data centres in
a local physical area,
such as a house, an office, a factory, or a small geographic area, where mobile
devices, such as 
drones, mobile robots, smartphones 
and fixed devices, such as sensors/actuators, workstations, and servers are
interconnected through single or multiple local area networks.

In this paper, we address the challenge of supporting FaaS in private edge cloud
systems. Off-the-shelf solutions to this challenge consist of deploying popular
open-source FaaS platforms (e.g., OpenFaas, Knative, Fission, OpenWhisk) on top
of container orchestration technologies (e.g., Kubernetes). However, these
technologies, which usually rely on containers and container orchestration
solutions, entail performance and resource overheads which can create issues on
devices with constrained resources---they might not have enough memory to host containers or computational power to effectively run functions in low-latency application contexts.

These problems motivated researchers and practitioners to consider alternatives
and propose runtimes that provide the isolation and parallel execution of
existing FaaS platforms yet mediate the heavy toll of the mentioned more complex
runtimes. Examples of these proposals include using virtual machines like that
of Java~\cite{SHMMC19} and Python~\cite{GFRC20} or embedding functions in
unikernels~\cite{MSKP20}. Unfortunately, while these solutions achieve the goal
of reducing the overhead of containers,
they respectively miss fundamental features. Java/Python VMs do not provide
high-performing runtimes~\cite{JPBG19} and properly isolate functions (e.g.,
exposing the users to security risks). Unikernels are still a niche technology
whose usage requires specific engineering knowledge (e.g., to define the minimal
OS stack needed to run high-level functions).

A promising alternative is leveraging
WebAssembly\footnote{\url{https://webassembly.org/}.} (Wasm) for lightweight
FaaS environments~\cite{KF23a} (introduced in more detail in
\cref{preliminaries}). Indeed, Wasm comes with a stack-based virtual machine
designed for running programs in a sandbox environment with performance close to
native code and fast load times. Wasm proved to be a valid candidate for FaaS,
providing lightweight sandboxing at the edge with both small latencies and
startup times~\cite{HR19,GMPCP20}---recently, providers like Cloudflare proposed
closed-source solutions based on
Wasm\footnote{\url{https://developers.cloudflare.com/workers/runtime-apis/webassembly/}.}.

\noindent\emph{FunLess}.\hspace{.5em} Building on these encouraging results, we
propose FunLess, a FaaS platform designed for private and hybrid edge-cloud
systems powered by a Wasm-based function execution engine. The advantages of
leveraging Wasm in FunLess are multifaceted:
\begin{itemize}
      \item \emph{Security.} Wasm's inherent security and isolation mechanisms make it
            well-suited for scenarios where data integrity and confidentiality are critical.


      \item
            \emph{Memory and CPU footprint.}
            FunLess does not require a container runtime (e.g., Docker) and orchestrator
            (e.g., Kubernetes). 
            Hence, the ``bare-metal'' deployment of FunLess frees resources essential for
            running functions on memory-constrained or low-power edge devices.
            %
      \item
            \emph{Cold starts.}
            FunLess leverages Wasm to mitigate the problem of cold starts~\cite{VFA20},
            i.e., delays in function execution due to the overhead of loading and
            initialising functions; an issue that constrained-resource edge devices can
            accentuate. Indeed, cold-start mitigations usually rely on caching and even
            keeping ``warm'' function instances and the memory footprint of containers can
            make these solutions unfeasible on constrained-resource devices. Contrarily, the
            small size of Wasm functions makes caching (and even fetch-and-load roundtrips)
            affordable in FunLess. This possibility, together with the fact that Wasm
            runtimes are optimised for fast startup times (note that Wasm's main use case is
            in-browser execution, where responsiveness is crucial) allow FunLess to achieve
            small cold-start overheads.

      \item
            \emph{Consistent function development and deployment environment.} Since Wasm
            abstracts away the hardware and environment it runs within, FunLess provides a
            consistent development and deployment experience across the diverse private edge
            architectures, offering a built-in solution for the challenges of variability in
            hardware and software environments of private edge-cloud scenarios. Similarly to
            Java bytecode, Wasm binaries can run on any platform that can execute a
            (dedicated) Wasm runtime. As illustrated in \cref{sec:architecture}, the
            developers only need to write once their functions\footnote{FunLess users can
                  write functions in any language supported by the platform, currently Rust, Go and JavaScript.},
            compile them into Wasm binaries, and load them into the platform. FunLess handles the task of
            running them on the possible diverse devices and architectures in the
            cloud/edge.

      \item
            \emph{Simple and flexible platform deployments.}
            FunLess can exploit existing containerisation solutions (e.g., Kubernetes) to
            streamline and ease its deployment. When container orchestration technologies
            are not affordable/available, users can install FunLess by running a
            \textit{Core} component (with metrics and storage services, e.g., resp.\@
            Prometheus and Postgres) on a node and a \textit{Worker} component on the nodes
            tasked to run the functions (cf.~\cref{sec:architecture}). This is both due
            to the use of WebAssembly (the binaries do not need an ulterior container for their isolation),
            and the built-in communication mechanism between nodes.

\end{itemize}




\noindent\emph{Evaluation of FunLess}.\hspace{.5em} Besides presenting FunLess,
in \cref{evaluation}, we report initial evaluation results obtained by
considering different deployment scenarios and by running several
typical cloud and edge FaaS workloads. Using these settings, we contrast the
specificity and originality of FunLess against existing FaaS platforms by
comparing it with three production-ready, widely adopted open-source FaaS
solutions: OpenFaaS, Fission, and Knative (we describe our selection process and
the peculiarities of these platforms in \cref{preliminaries}). The first
observation we highlight is that FunLess is the only platform we could deploy on
an edge-only cluster of constrained devices (Raspberry Pi 3B+)---the other
platforms can use the edge devices to run functions but the platforms themselves
cannot entirely run on these devices because their components need more powerful
machines.
To consolidate the above observation, we profile the memory footprint of the
different platforms. 
As expected, the deployment that consumes the minimum amount of memory is that
of FunLess in bare-metal modality, i.e., without Kubernetes' support. Another
important qualitative result is that only FunLess and Fission support the
seamless integration of heterogeneous devices---the other platforms require
ad-hoc compilations of the functions for different target devices. This result,
combined with the one about memory footprint, supports our claim that FunLess is
the only alternative that is fully deployable on heterogeneous clusters of
constrained edge devices.
The performance of FunLess is generally
comparable with the other platforms (that use binary native code). Indeed, considering
the average latency, FunLess in the cloud-edge scenario has a performance gap of
at most 5\% from the best-performing solution, with the exclusion of the
compute-intensive test case, where using native code makes a more significant
difference. We do not see
these latter results as negative (indeed, we were expecting an even higher
performance gap) and they rather compel an interesting case for investigating a
problem orthogonal to FunLess, since they witness how using different languages
(and language runtimes) for function implementation influences their
performance. Nonetheless, we expect that including improved versions of Wasm
engines in FunLess could greatly increase the performance of all cases,
indicating not only that FunLess's approach is feasible but that it is a
scalable solution able to integrate improvements from the Wasm community.
Besides these observations, in
\cref{sec:related_work}, we compare FunLess with the existing alternatives from
the literature (the closest, based on Wasm), commenting on the relevant traits
that distinguish our proposal and concluding in \cref{sec:conclusion} by discussing directions for future work.

\section{Preliminaries}
\label{preliminaries}

We dedicate this section to providing the preliminary notions useful to
contextualise our contribution. Specifically, we introduce WebAssembly---the
technology underpinning the FunLess execution runtime (cf.\@
\cref{sec:architecture})---and briefly present the serverless platforms we
compare against in our evaluation (cf.\@ \cref{evaluation}).

\subsection{WebAssembly}

The WebAssembly~\cite{web:webassembly} technology, Wasm for short, is a W3C
standard since 2019, maintained with contributions from Apple, Google,
Microsoft, Mozilla, and other companies.

The idea behind Wasm is to provide a simple assembly-like instruction set which
can run efficiently within a browser. At its core, Wasm includes a binary
instruction format 
and a stack-based Virtual Machine that supports functions and control flow
abstractions like loops and conditionals.

Although browsers are the main target of Wasm, recent initiatives, like
WebAssembly System Interface~\cite{web:wasi} (WASI) normed the implementation of
Wasm runtimes that support the execution of Wasm code outside the browser with a
set of APIs that provide POSIX capabilities (e.g., file system, network, and process management).
Some examples of WASI-compliant runtimes, either open-source or
proprietary, are Wasmtime~\cite{web:wasmtime}, Wasmer~\cite{web:wasmer}, and
WasmEdge~\cite{web:wasmedge}.

Focussing on FaaS, Wasm provides a sandboxed runtime environment for functions,
akin to containers. However, while one needs to build a container (for the same
function) for each targeted architecture, the same Wasm binary can run on
different architectures thanks to the hardware abstraction provided by the Wasm
runtime. Moreover, Wasm binaries tend to be more lightweight than containers,
thanks to the fact that they do not need to include a pre-packaged filesystem.





\subsection{Alternative Open-Source Serverless Platforms}
\label{sec:platforms}
As part of our platform's performance assessment, we select and characterise
three popular alternative open-source serverless solutions we compare FunLess
against.

We aimed to select widely adopted open-source solutions for serverless. These
are not necessarily adapted for the private edge cloud case since, given the
recent emergence of this case, there are no popular open-source proposals yet.
    
    To select the candidates, we searched GitHub for the keyword ``faas''
    (which, at the time of writing returned 3.9k matches) and we followed four
    inclusion criteria for the selection: production-ready (used in industry,
    verified by looking at the commercial testimonials found on the project's
    webpages), popular (above 5k stars on GitHub), actively developed (commits
    within the last quarter and with at least 100 contributors), and able to run
    on both AMD64 and ARM hardware (i.e., the most common hardware found in the
    cloud and edge devices). We ranked the results by popularity (GitHub stars)
    and selected the first three. The selected platforms, in ranking order, are
    OpenFaaS (24k+ stars), Fission (8k+ stars), and Knative (5k+
    stars).\footnote{Resp.\@ found at \url{https://github.com/OpenFaaS/faas},
    \url{https://github.com/fission/fission}, and
    \url{https://github.com/knative}. Note that the above selection lacks Apache
    OpenWhisk, a popular ({6k+ stars}), serverless platform which we excluded
    because the container images of its main components (Controller and Invoker)
    are not available for ARM architectures, thus precluding its installation on
    many edge devices.}

We conclude this section with the main traits of the
alternatives.

\paragraph{OpenFaaS}
OpenFaaS builds on Kubernetes, and it takes advantage of Kubernetes' scheduler
for function allocation and scaling---specifically, functions are pods, i.e.,
application containers that enclose the function's code and runtime environment.
Since pods are generic containers, OpenFaaS supports different languages by
providing language-specific template containers with example source files that
users can extend to implement their functions and include the necessary
dependencies.

\paragraph{Fission}
Like OpenFaaS, also Fission builds on Kubernetes. However, Fission does not rely
on user-built containers for functions, allowing users to directly upload their
source code. Functions run through the use of ``environments'', which
essentially define which pre-built containers the platforms shall use to run the
functions' sources (as compiled binaries or via an interpreter).
One of the strengths of Fission is the small cold-start times it affords for
functions deployed as source code (i.e., not binary executables). To achieve
this result, Fission initialises ``general-purpose'' containers for the language
environment of the deployed functions (e.g., if the user deploys NodeJS
functions, Fission prepares a pool of NodeJS containers at each execution
machine). At function invocation, Fission uses one of these ``warm'' containers
by injecting and running therein the code of the function. 

\paragraph{Knative}
Knative is also a Kubernetes-based serverless platform. The main difference with
OpenFaaS and Fission is that Knative adopts a more low-level approach to
function development. Essentially, developers implement their functions as
containerised microservices (the main contemporary programming style
complementary to serverless for cloud-based distributed
software~\cite{DGLMMMS17}), which Knative executes in a serverless-like fashion
(managing event-based allocation and scaling).

\section{Platform Architecture}
\label{sec:architecture}

We present the principles and technologies behind FunLess, its architecture and
discuss our design choices (trade-offs and limitations).

\begin{figure*}[h]
    \centering
    \includegraphics[width=0.95\textwidth]{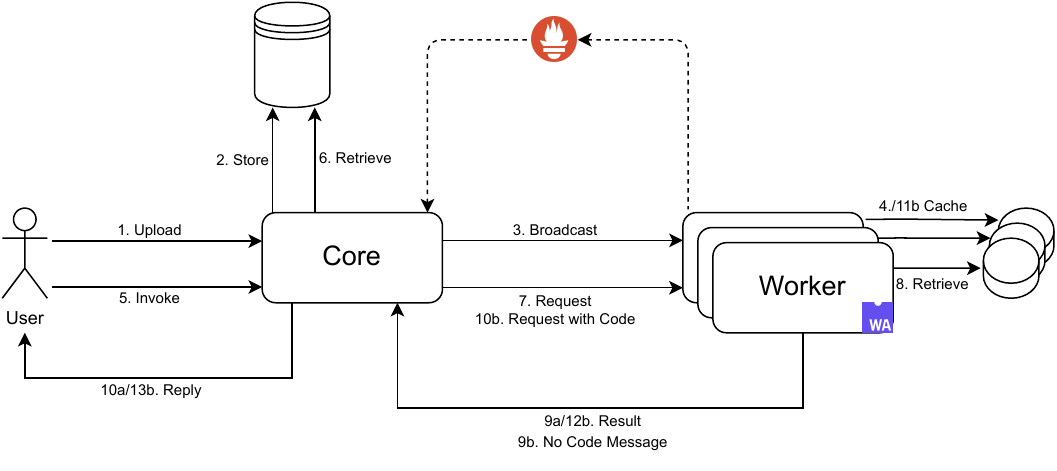}
    \caption{Architecture of the FunLess platform with the function flow from
        creation to invocation.}
    \label{fig:architecture}

\end{figure*}

The main principles behind the design of FunLess are the simplicity of both
function development and platform deployment and the flexibility of hardware and
deployment automation. In particular, FunLess is independent of the underlying
deployment orchestrators (if any), which avoids potential overheads and allows
users to install the entire platform on resource-constrained, low-power edge
devices. For the implementation of the platform, we used
Elixir~\cite{J24}, which is a functional,
concurrent, high-level general-purpose programming language that runs on the
BEAM virtual machine~\cite{beam} (used by the Erlang language).
Specifically, Elixir and the BEAM allowed us to simplify the creation and
deployment of a distributed application without relying on container
orchestration technologies, while retaining high performance, fault-tolerance,
and resilience (provided by the BEAM's scheduler and lightweight processes,
famous for being optimised for concurrent and distributed systems).

We represent in Figure~\ref{fig:architecture} both the components that make up
the platform's architecture and the typical flow developers and users follow to
create and invoke functions. Architecture-wise, FunLess consists of mainly two
components: the \textit{Core} and the \textit{Worker}, which we detail in the
next parts of this section. Briefly, the \emph{Core} acts as an user-facing API
to \emph{i}) create, fetch, update, and delete functions and \emph{ii}) schedule
functions on workers. The \emph{Worker} is the component deployed on every node
tasked to run the functions; in the remainder, we refer to these nodes as
\textit{Worker}s. In addition to these two components, FunLess includes a
\textit{Postgres} database to store functions and metadata and
\textit{Prometheus} to collect metrics from the platform.\footnote{Resp.\@ found
    at \url{https://www.postgresql.org/} and \url{https://prometheus.io}.}

FunLess is an open-source project and both its source code~\cite{web:funless-repository}
and documentation~\cite{web:funless-website} are publicly available.

\subsection{Core}

The \textit{Core} is the controller of the platform. It exposes an HTTP REST API
to the users, handles authentication and authorization, and manages functions'
lifecycle and invocations.

Although the \textit{Core} implements the main coordination logic and
functionalities of FunLess, it is a lightweight component. For instance, on a
Raspberry Pi 3B+ its local bare-metal deployment (that includes the database,
the monitoring system and the underlying operating system and services) occupies
ca.\@ 600 MB of RAM when idle.



Functionality-wise, FunLess users create a new function by compiling its
source code to Wasm---using either the language's default compiler (for Rust),
an alternative one (for Go), or an external tool (for JavaScript)---and
uploading the resulting binary to the \textit{Core}, assigning to it a name.
Users can group functions in modules and, when uploading a function, they can
optionally specify which module the function belongs to. Moreover, users should also specify the amount of memory reserved for the execution of the function.

Looking at the steps reported in \cref{fig:architecture}, once the \textit{Core}
receives the request to create a function (1. Upload), it stores its binary in
the database (2. Store). Fetch, update, and deletion happen via the assigned
function name. When the \textit{Core} successfully creates a function, it
notifies the \textit{Worker}s (3. Broadcast) to store a local copy of the
function binary (4. Cache) compiled from the source code with the given metadata
(i.e., module and function names). This push strategy helps to reduce part of
the overhead of cold starts. Indeed, most FaaS platforms follow a pull policy
where, if the execution nodes do not have the function in their cache (e.g., it
is the first time they execute), they fetch, cache, and load the code of the
function, undergoing latency. The small occupancy of Wasm binaries makes it
affordable for FunLess to employ a push strategy, helping to reduce cold-start
overheads.


Since both the \textit{Core} and the \textit{Worker}s run on the BEAM, these
components communicate via the BEAM's built-in lightweight distributed
inter-process messaging system, avoiding the need (complexity, weight) for
additional dependencies for data formatting, transmission, and component
connection.

When a function invocation reaches the \textit{Core} (5. Invoke), the latter
checks the existence of the function in the database and retrieves its code (6.
Retrieve). If the function is present in the database, the \textit{Core} uses
the most recent metrics---we represent the pushing of the data, updated every 5s
by default, from Prometheus to the Core with the dashed line in
\cref{fig:architecture}---to select on which of the available \textit{Worker}s
to allocate the function (7. Request). The selection algorithm starts from the
\emph{Worker} with the largest amount of free memory to the one with the
smaller. If no worker has enough memory to host the function, the invocation will return with an error.

%
%
After the \textit{Worker} successfully ran the function (we detail this part of
the workflow in the section about \textit{Worker}s, below) it sends back to the
\textit{Core} the result (if any), which the \textit{Core} relays back to the
user (10a/13b. Reply). If no \textit{Worker} is available at scheduling time or
there are errors during the execution, the \textit{Core} returns an
error.


Another important feature of FunLess is that the \emph{Core} can automatically
discover the \emph{Worker}s in its same network. This feature derives from
Elixir's libcluster
library\footnote{\url{https://hexdocs.pm/libcluster/readme.html}.}, which
provides a mechanism for automatically forming clusters of BEAM/Erlang nodes.
Technically, when deployed on bare metal, FunLess follows the Multicast UDP
Gossip algorithm of the library, to automatically find available workers.
Instead, when deployed using Kubernetes, FunLess relies on the service discovery
capabilities of the container orchestration engine to connect the \emph{Core}
with the \emph{Worker}s, paired with the ``Kubernetes'' modality of the library.
Users can manually connect \emph{Worker}s from other networks via a simple
message (e.g., a ping) thanks to the BEAM's built-in capability of connecting to
other BEAM nodes.

\subsection{Worker}
The \textit{Worker} executes the functions requested by the \textit{Core}. The
\textit{Worker}s employs Wasmtime, a standalone runtime for Wasm and WASI by the
Bytecode Alliance~\cite{web:bytecodealliance}. The main reasons behind this
choice come from the ease of integration, amount of contributors, and
security-oriented focus of the project.
While \textit{Worker}s integrate Wasmtime, we modelled their architecture to
abstract away the peculiarities of specific Wasm runtimes so that future
variants can use different runtimes and even extend support for multiple ones
(possibly letting users specify which one to use).

When a \textit{Worker} receives a request from the \textit{Core} to execute a
function (7. Request), it first checks whether it has a cached version of the
function's binary (8. Retrieve). If that is the case, it loads and runs the
function's binary and returns to the \textit{Core} the result of the computation
(9a. Result). If the \textit{Worker} does not find the code of the function in
its local cache, it contacts the \textit{Core} (9b. No Code Message), which
responds with a request that carries the code of the function to the
\textit{Worker} (10b. Request with Code). Upon reception, the \textit{Worker}
compiles the code, caches the binary for future invocations (11b. Cache), loads
it to run the function, and relays the result to the \textit{Core} (12b.
Result).

The above mechanism is an important advantage afforded by FunLess for the edge
case. Function fetching (if needed) transmits small pieces of binary code
(rather than heavyweight containers). Wasm binaries achieve the two-fold
objective of having \textit{Worker}s run functions on different hardware
architectures (e.g., AMD64, ARM) and allowing users to write their functions
once, knowing that they will execute irrespective of the hardware of the
\textit{Worker}.

Summarising, fetching and precompiling (if any, depending on cache status)
constitutes most of the ``cold start'' overhead in FunLess, which the platform
greatly reduces w.r.t.\@ alternatives relying on containers (which are heavier
both in terms of bandwidth and memory occupancy).



Regarding caching and eviction, \textit{Worker}s set a threshold for the cache
memory (configurable at deployment time). If the storing of a new function
exceeds that threshold, the \textit{Worker} evicts the function with the longest
period of inactivity (invocation- or update-wise). Additionally,
\textit{Worker}s automatically evict functions if inactive for a set amount of
time (by default, 45 minutes).

\subsection{Design choices and limitations}

Since a small resource footprint and simplicity (of architecture and
computation) are the driving principles behind FunLess' implementation, we
favoured design choices (both w.r.t.\@ the components in the architecture and
the internal implementation) that introduced the least complexity while
affording flexibility (of implementation and deployment). Below, we discuss the
main aspects that FunLess trades off for the above benefits.

\noindent\emph{Language support}.\hspace{.5em} FunLess requires functions to be
compiled to Wasm to execute them. Moreover, for the Wasm binary to properly
integrate with the \textit{Worker}, it needs to expose a specific function that
acts as a ``wrapper'' for the user's function. The wrapper performs input and
output (de)serialisation, and is not a standard feature of Wasm modules.
Therefore, FunLess provides a wrapper for each supported language---depending on
the language, a wrapper can be a library, macro or compiler extension. While
offering support for different languages is not essential for this presentation,
FunLess already supports three languages: Rust, Go and JavaScript---and we
planned support for more in the future. Specifically, we chose Rust for its
performance, its growing developer community, and its ease of compiling to Wasm;
similarly, Go is famous for its performance and widespread use in cloud
computing; lastly, JavaScript is one of the most popular languages in software
development.

\noindent\emph{Resilience}.\hspace{.5em} FunLess's \textit{Core} component,
which acts as the sole scheduler and holder of the platform's state, is not
replicated. On the one hand, this reduces the footprint of the platform since
users just need to deploy one \textit{Core}.
On the other hand, the \textit{Core} is a single point of failure of the
architecture. The BEAM opportunely guarantees fault-tolerance, so that the
\textit{Core} can recover from software crashes. However, the platform would
stop working properly if the hardware hosting the \textit{Core} failed. On
software crashes, the only data lost are the invocations in transit (which the
users would notice as timed out), but the rest of the system would recover
(normal functionality, connections to the \textit{Worker}s, metrics, and
storage), following the connection protocols mentioned above.

\noindent\emph{Robustness}.\hspace{.5em} FunLess implements an at-most-once
message relay policy, hence, lost messages between the \textit{Core} and
\textit{Worker}s imply the failure of the invocation. Implementing more robust
semantics, e.g., at least once, would require the inclusion of a message broker,
increasing the load on nodes and the architecture's complexity.

\noindent\emph{Retry policies}.\hspace{.5em}
The \textit{Core} does not implement retry policies. Thus, if a function's
execution fails on the chosen \textit{Worker} or that \textit{Worker} becomes
unresponsive, the \textit{Core} does not try to run the function on another
worker. Implementing retry policies would increase the complexity platform-wide.
Specifically, the \textit{Core} would need to keep track of the state of
function invocations, increasing the amount of coordination/messages with the
\emph{Worker}s. This extension would also increase the amount of data and
interactions with the database (needed to enforce the transactional management
of functions' state and stave off the risk of losing this data due to crashes)
and further complicate the \textit{Core}'s implementation to manage back-off
strategies and execution time limits. Nonetheless, we plan to implement retries
with an ``opt-in'' approach (the BEAM already provides some building blocks for
the task, used to implement function timeouts and monitoring), giving users the
flexibility to choose between a lighter setup or increased reliability.

\section{Empirical Evaluation}
\label{evaluation}

We now present our evaluation of the FunLess platform. We aim to verify the
platform's capability to run without underlying orchestrators or container
engines, as well as the possibility of deploying the entire stack (including
database and monitoring system, cf.\@ \cref{sec:architecture}) on
resource-constrained edge devices.

\noindent\emph{Test Infrastructure}.\hspace{.5em} As mentioned in
\cref{preliminaries}, we compare FunLess with OpenFaaS, Fission, and Knative. To
structure the comparison, we assemble the following cloud and edge resources. As
edge devices, we use two Raspberry Pi 3B+ (1.4GHz quad-core CPU and 1 GB of
RAM), both with Debian 12 (Bookworm) ARM64 OS. For the cloud instances, we
use Terraform to programmatically provision the Virtual Machines (VMs) and
Ansible to install and configure the platforms.\footnote{Resp.\@ found at
    \url{https://www.terraform.io/} and \url{https://www.ansible.com/}.}
Specifically, we provision up to five VMs in the GCP cloud (Europe West region
with Ubuntu 22.04) and set a virtual private network up, to which we add the 2
nodes from the private edge (the Raspberry Pi devices). One e2-medium cloud node
(2 vCPUs, 4 GB memory) is for the Kubernetes control plane, and it is not used
to perform any other form of computation. In the cloud, we deploy the
\emph{Core} on an e2-medium node while the \emph{Worker}s are on n1-standard-1
(1 vCPU, 3.75 GB memory) VMs.

\noindent\emph{Deployment Configurations}.\hspace{.5em} On the above
infrastructure, we define four configurations for the deployment of the
platforms:

\begin{itemize}
    \item \emph{edge-only}: we only use the edge devices, one hosts the
          core/controller of the platform and one acts as a worker, without
          using Kubernetes;
    \item \emph{cloud-bare-edge}: the core/controller of the platform is in a
          cloud node and the two edge devices act as workers, without using
          Kubernetes;
    \item \emph{cloud-edge}: same as \emph{cloud-bare-edge}, but using
          Kubernetes;
    \item \emph{cloud-only}: we only use cloud nodes, one hosts the
          core/controller and three act as workers, using Kubernetes.
\end{itemize}

\subsection{Benchmarks}
\label{sec:setup}

The first result we report is that, due to the resource constraints of the edge
nodes, among all the platforms considered, only FunLess could be deployed in the \emph{edge-only} scenario. 
The reason
behind the result is that all alternatives need Kubernetes to run, and it has
too big of a memory footprint to run on the edge devices. Hence, in the
remainder, we limit the experiments with the alternative platforms to the other
three scenarios.

To test all configurations, we collect the latencies of all platforms using the
same set of microbenchmarks, drawn from the Serverless Benchmark Suite
(SeBS)~\cite{copik2021sebs}, including an additional compute-intensive benchmark
(i.e., matrix multiplication), inspired by Gackstatter et al.~\cite{GFD22}. To
measure the memory footprint of the platform, we use a simple ``hello world''
function (described later).


We use Go and JavaScript (JS) for the implementation of the functions since
these are the only two programming languages officially supported by all the
platforms under study.

The functions of the microservices are (the first three from SeBS):
\begin{itemize}
    \item \textit{sleep}, implemented in JS, it waits 3 seconds and returns a
          fixed response (``Slept for 3 seconds''). This benchmark tests a
          platform's capability of handling multiple functions running for
          several seconds and its requests queuing-management process.
    \item \textit{network-benchmark}, implemented in Go, it sends 16 HTTP
          requests with a timestamp and uploads this information to a cloud
          bucket. This test tracks how long each HTTP request takes to complete.
    \item \textit{server-reply}, implemented in Go, sends a message to a server
          and waits for a reply. Differently from \textit{network-benchmark},
          this test measures the performance of the network stack and the
          latency of the platform considering the entire duration of the
          function execution.
    \item \textit{matrixMult}, implemented in JS, multiplies two \(10^2\) square
          matrices and returns the result to the caller. This test measures the
          performance of handling compute-intensive functions.
\end{itemize}

For each platform, we define and build the functions following the approach
suggested by their respective documentation. For OpenFaaS, we write both Go and
JS functions using the platform's native templates, generating the related
container images. For Fission, we use the suggested Go and JS environments
(specifically, Fission uses a NodeJS runtime for JS and a builder for Go to
internally create the containers).
For Knative, we build the containers for Go using
ko\footnote{\url{https://ko.build/}.} and use a custom Dockerfile for JS due to
issues of Knative's integrated
tool\footnote{\url{https://github.com/knative/func}.} in building the container
for ARM64 (we build the container by mimicking the one used internally by
Knative's tool).
For FunLess, we compile JS using a customised version of the
javy\footnote{\url{https://github.com/bytecodealliance/javy/}.} project and Go
using TinyGo\footnote{\url{https://tinygo.org/}.}. Both compilers create a
binary with a language-specific wrapper that performs input and output parsing
at function invocation, simplifying the interaction with the \textit{Worker}
component (FunLess relies on a slightly modified version of javy that supports
this feature).

We invoke the functions and collect the metrics using Apache
JMeter\footnote{\url{https://jmeter.apache.org/}.}. Specifically, we record the
latency between the delivery of the request of the function invocation and the
reception of the response. We set up JMeter to invoke in parallel the functions,
except for \textit{network-benchmark}, where we use only 1 thread to avoid
concurrency issues due to parallel accesses to the bucket.

For each benchmark, we aim to collect 800 data points, with the only exception
of \textit{sleep}, where we stopped at 100 repetitions due to the predictable
duration of this function. In summary, for \textit{server-reply} and
\textit{matrixMult}, we run 4 parallel threads, each sending 200 sequential
requests (800 invocations in total); for \textit{network-benchmark}, we use 1
thread performing 50 iterations, each sending 16 HTTPS requests (800
HTTPS request in total); for \textit{sleep}, we use 4 parallel threads, each
sending 25 sequential requests (100 invocations in total). We repeat each benchmark 5 times.

In addition to latency, 
we track 
the memory usage of the various FaaS components
on the edge devices by invoking a simple
function to measure the baseline for the memory footprint and what happens when
a small (memory-wise) function runs. Namely, we write an ``hello world'' JS
function, dubbed \textit{hellojs}, which parses the input parameters and returns
a string. We invoke the function continuously on 4 parallel threads for 5 minutes
(i.e. starting a new request as soon as the last one has completed, on each thread),
keeping a sampling rate of one second to track the memory used by the
edge devices. When measuring the memory footprint, we use a custom deployment
with only one node (a Raspberry Pi) running the functions, and the rest of the
components (e.g., schedulers, databases, web servers, etc.) deployed in the
cloud. Finally, for the edge-only deployment of FunLess, we measure the memory
usage of the \textit{Core} on a second Raspberry Pi.

\begin{figure*}[h]
    \centering
    \includegraphics[width=.49\textwidth]{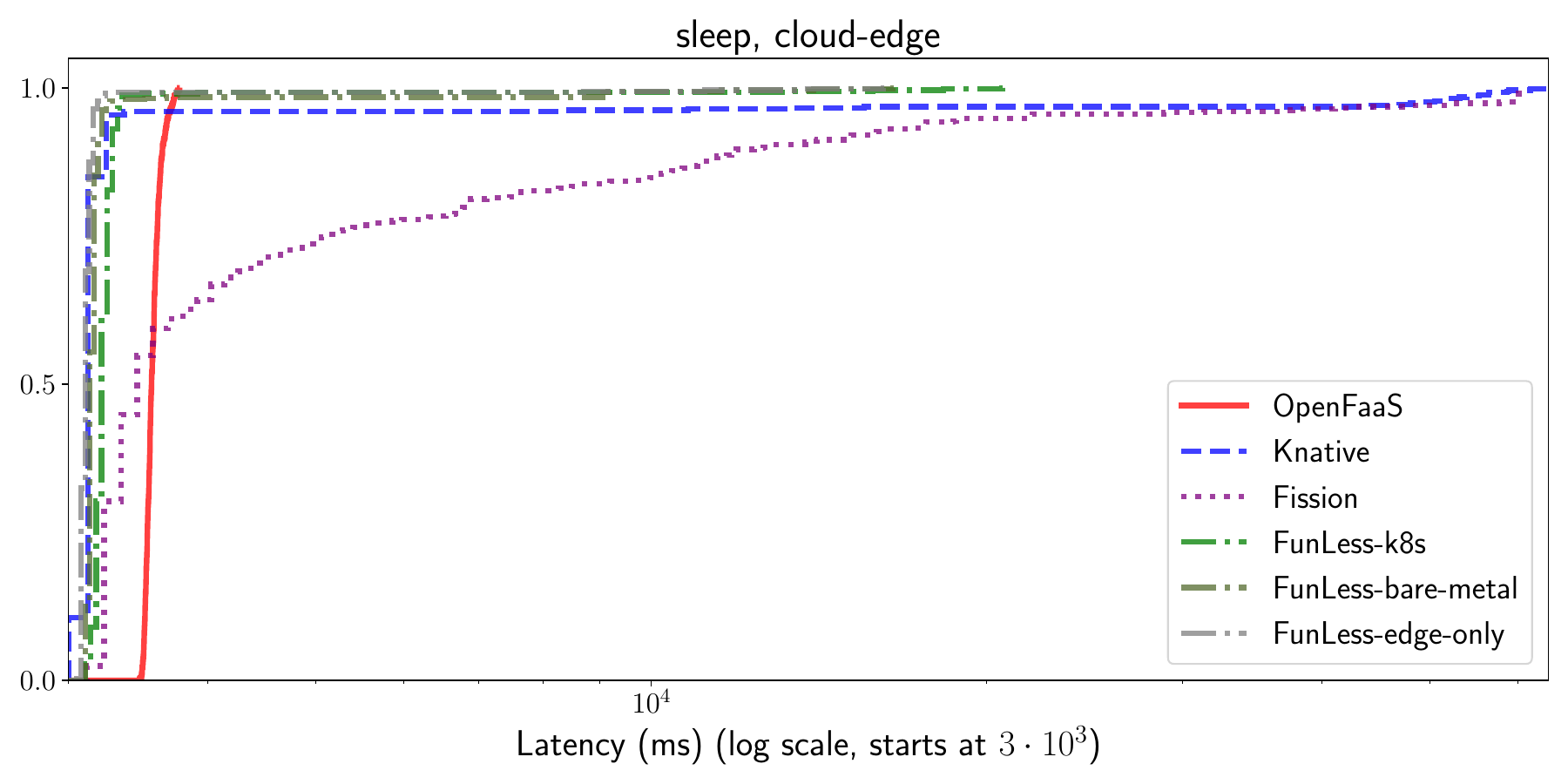}
    \includegraphics[width=.49\textwidth]{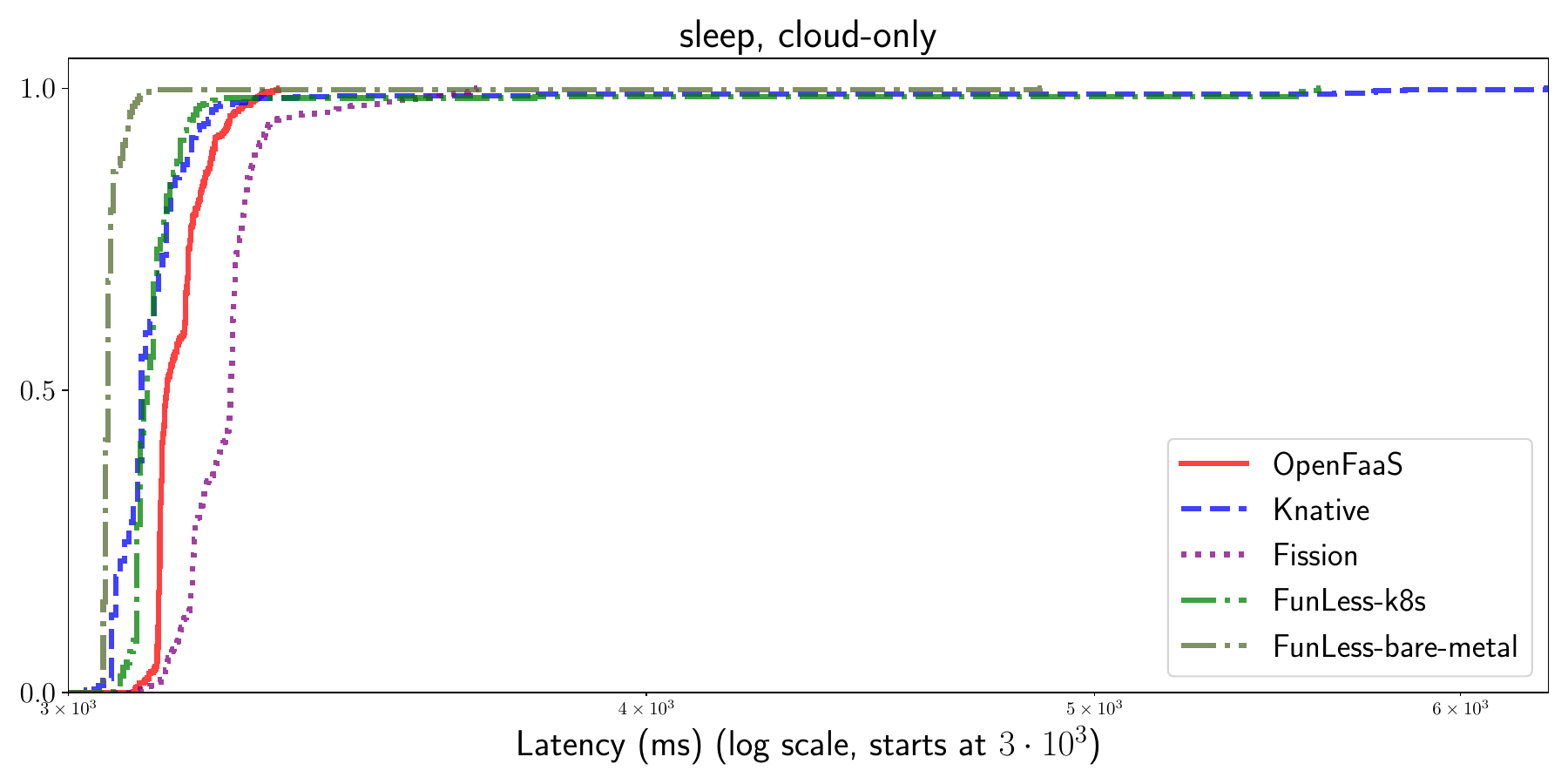}
    \includegraphics[width=.49\textwidth]{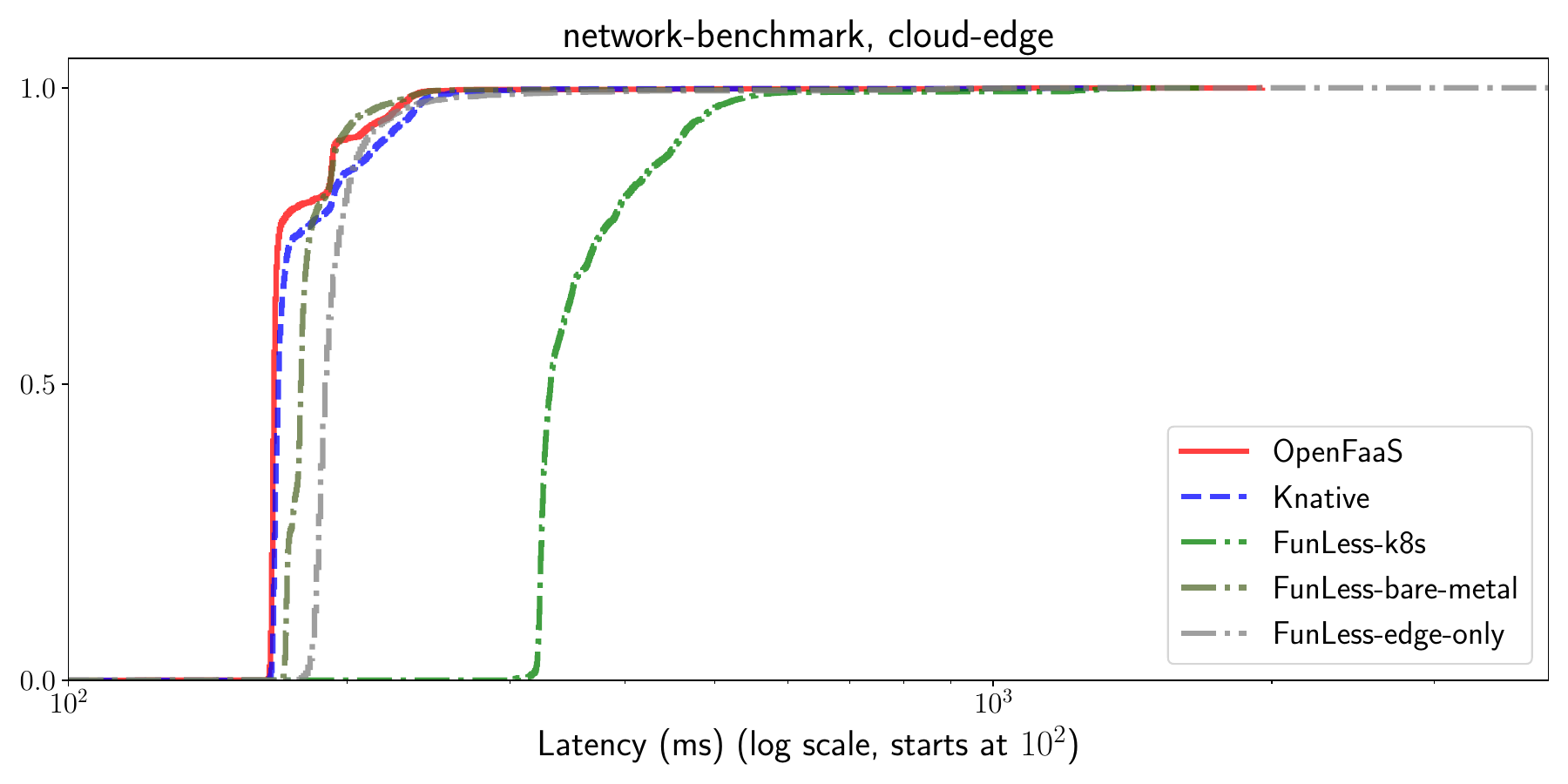}
    \includegraphics[width=.49\textwidth]{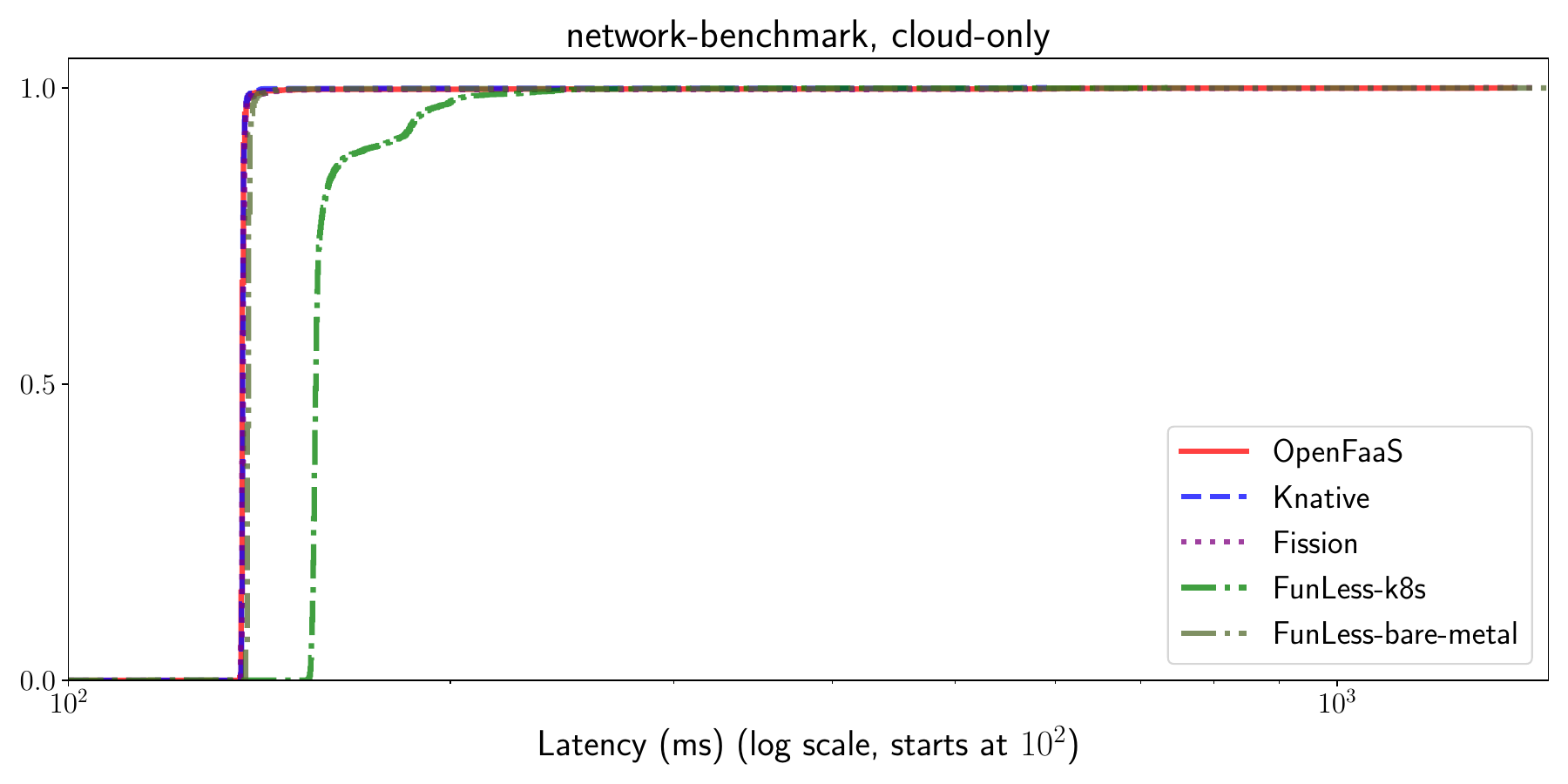}
    \includegraphics[width=.49\textwidth]{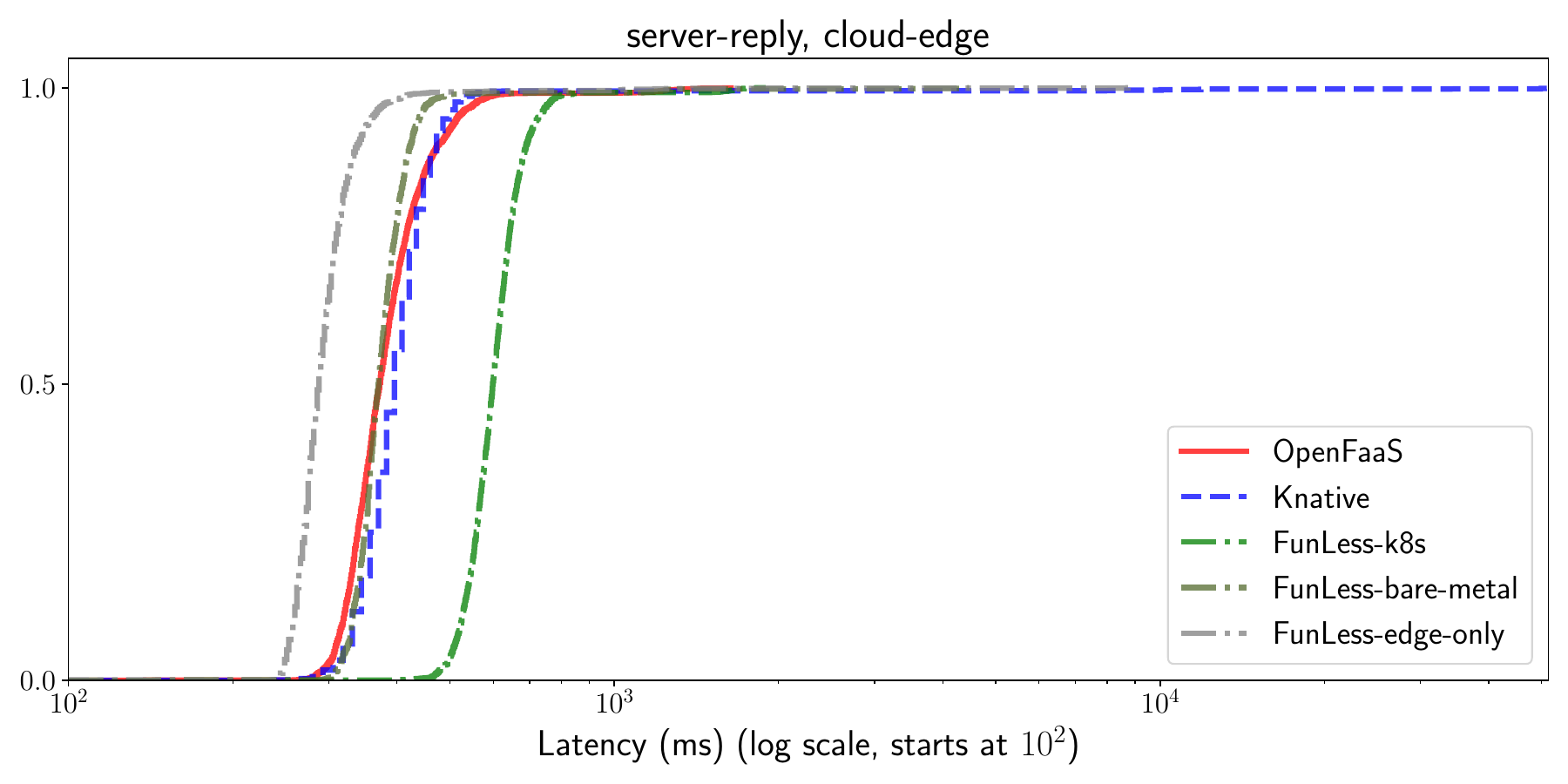}
    \includegraphics[width=.49\textwidth]{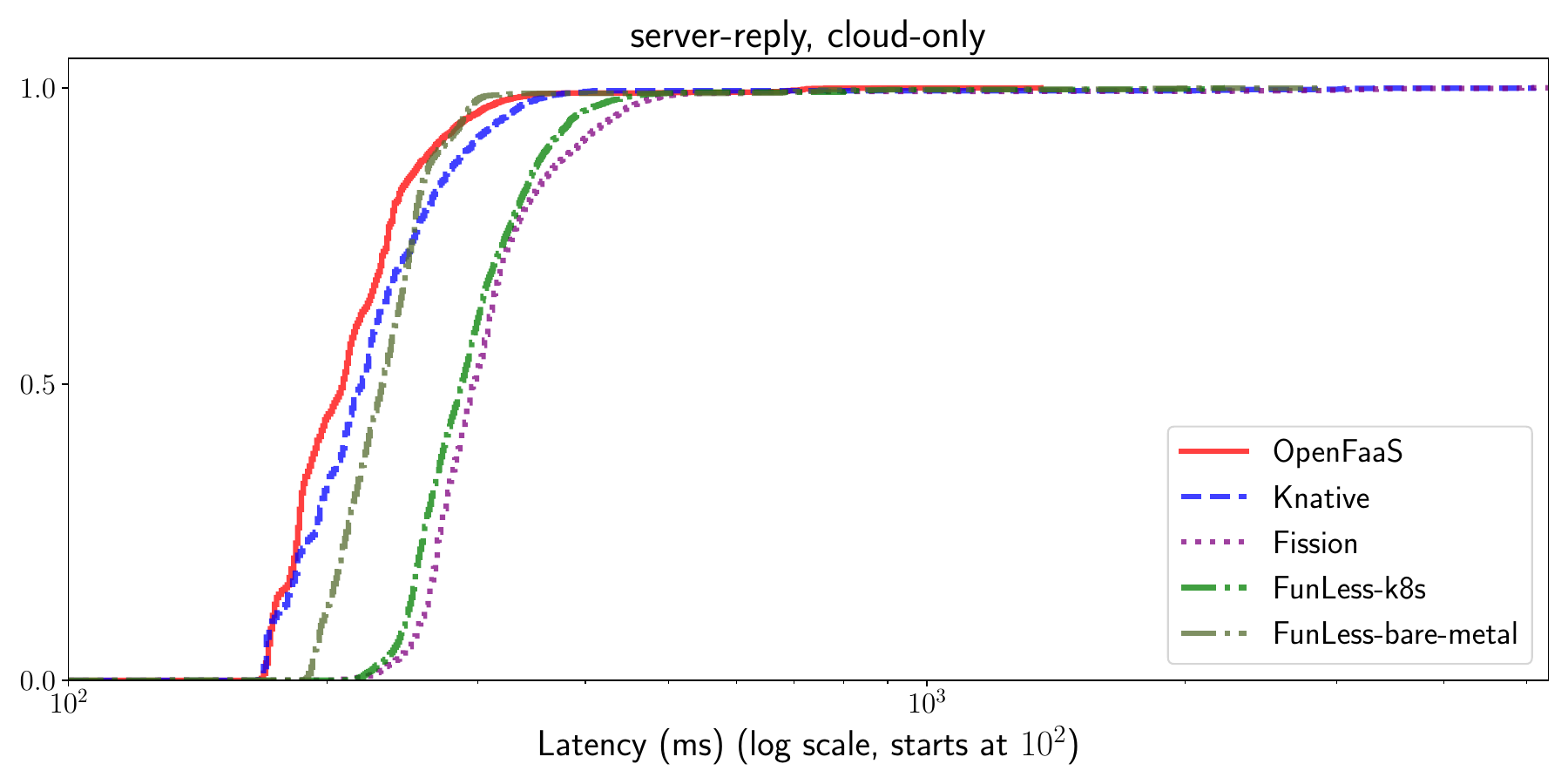}
    \caption{Cumulative distribution of the latencies of, from top to bottom,
        resp.\@ the \textit{sleep}, \textit{network-benchmark}, and
        \textit{server-reply} benchmarks. At each row, on the left, we show the
        \textit{cloud-edge} scenario and, on the right, the \textit{cloud-only}
        one.}
    \label{fig:results}
\end{figure*}

\subsection{Results}
\label{sec:results}

We show the results of the different benchmarks and scenarios in
\cref{fig:results}, reported as cumulative latency distributions---from top to
bottom, the rows report resp.\@ the \textit{sleep}, \textit{network-benchmark},
and \textit{server-reply} and, at each row, on the left we show the
\textit{cloud-edge} scenario and on the right the \textit{cloud-only} one.

Since FunLess is the only platform able to run in the \textit{edge-only} and
\textit{cloud-bare-edge} scenarios, for compactness, we report these results in
the \textit{cloud-edge} plots (left column of \cref{fig:results}), resp.\@ with
the lines called \textsf{FunLess-edge-only} and \textsf{FunLess-bare-metal}; the
other line for FunLess in the plots is \textsf{FunLess-k8s}, which corresponds
to Funless deployed using Kubernetes.

We start from the first-row plots in \cref{fig:results}, relative to
\textit{sleep}. In the \textit{cloud-edge} case, FunLess (the different
modalities differ very little) and Knative are the best-performing platforms,
although Knative has a few long-running outliers. Then, we find OpenFaaS with a
narrow distribution but generally much slower than FunLess and Knative. The
worst-performing platform is Fission, which presents a few good data points
(closer to the origin), followed by a widely spread distribution of slow
instances. In the \textit{cloud-edge} case, the performance differences among
the platforms narrow down. We interpret this fact as an indication that OpenFaaS
and Fission need more powerful machines to run properly. Interestingly, in this
scenario, \textsf{FunLess-bare-metal} has the best performance, followed by its
Kubernetes variant and Knative. While the differences are small, we attribute
this result to the higher load exerted by Kubernetes on the infrastructure.



Looking at the second row of \cref{fig:results}, that of
\textit{network-benchmark},
in both \textit{cloud-edge} and \textit{cloud-only} scenarios, all platforms
except FunLess perform similarly, with Knative and OpenFaaS scoring the best
results on average. Note that the plot line for Fission is missing in the
\textit{cloud-edge} case, due to memory issues that prevent the platform from
running this benchmark\footnote{The issue comes from Fission's Go builder and
    containers pool which, together, exceed the available capacity of the Raspberry
    Pis and go out of memory.}.
%
Notably, \textsf{FunLess-k8s} is the worst-performing platform, while its
bare-metal variant aligns with the performance of the alternatives. We attribute
this result to the overhead due to the interplay between Kubernetes' network
stack and the current implementation of the FunLess Wasm runtime. Specifically,
the current version of Wasm runtime equipped by FunLess (Wasmtime 12.0.1)
does not support natively HTTP requests. To enable this
functionality, FunLess' \textit{Worker}s implement auxiliary operations that the
Wasm functions can invoke to issue an HTTP request and obtain a response. Using
these operations means that the Wasm function yields control to the BEAM, which
executes the HTTP request and then relays the response (and control) back to
Wasm. In the configuration with Kubernetes, this back-and-forth between the BEAM
and Wasm is further weighed down by Docker's and Kubernetes' network stack,
resulting in higher latencies.


The bottom benchmarks in \cref{fig:results}, \textit{server-reply}, shows
similar results to \textit{network-benchmark}. Indeed, \textsf{FunLess-k8s}
performs worse than most of the alternatives, although Fission is the
worst-performing one in the \textit{cloud-only} scenario.
As explained above, this degradation derives from overheads generated by HTTP
requests and the interplay among the current implementation of the Wasm
runtime, the BEAM, Docker, and Kubernetes.

\begin{figure*}[t]
    \centering
    \includegraphics[width=.49\textwidth]{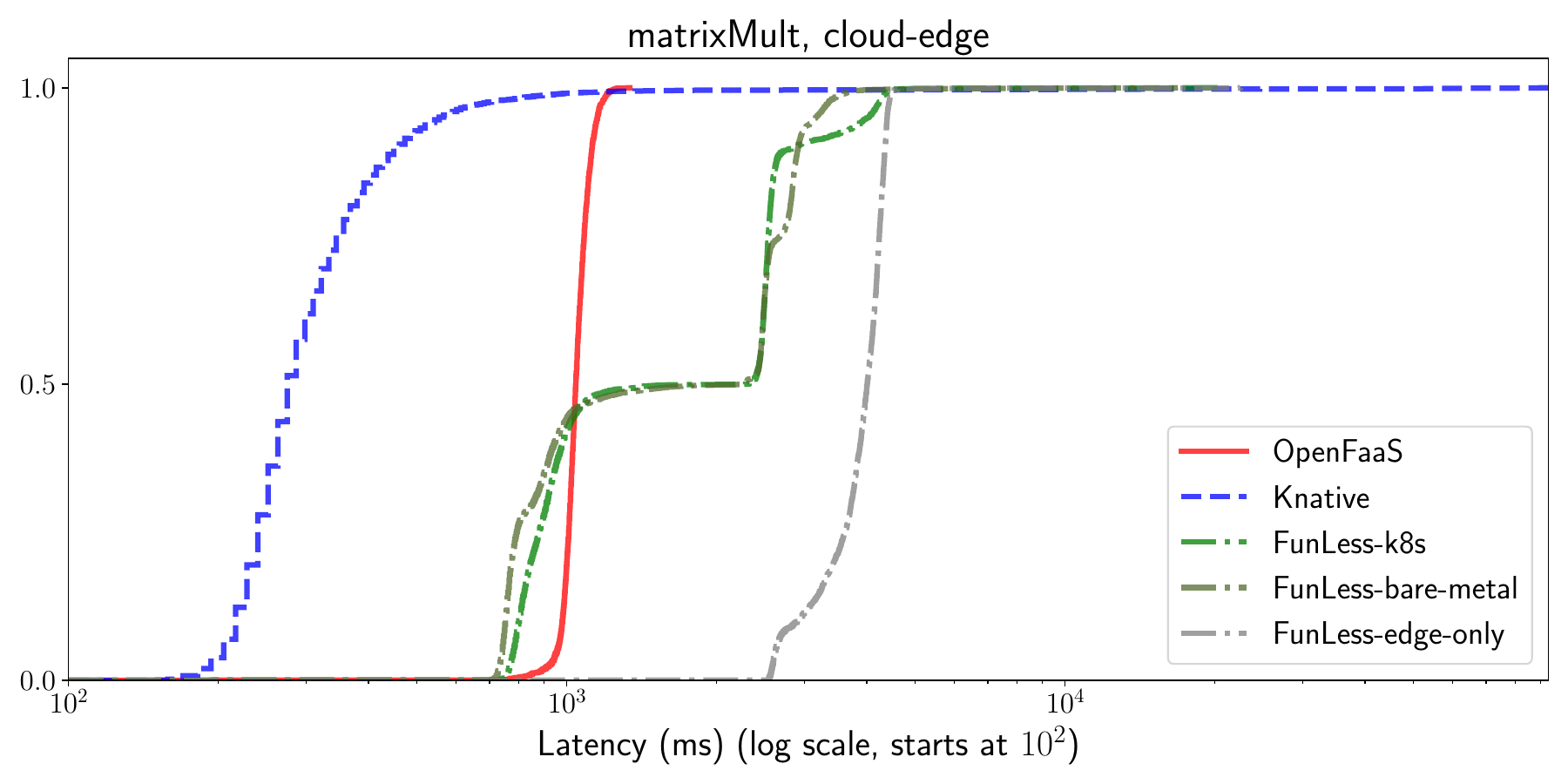}
    \includegraphics[width=.49\textwidth]{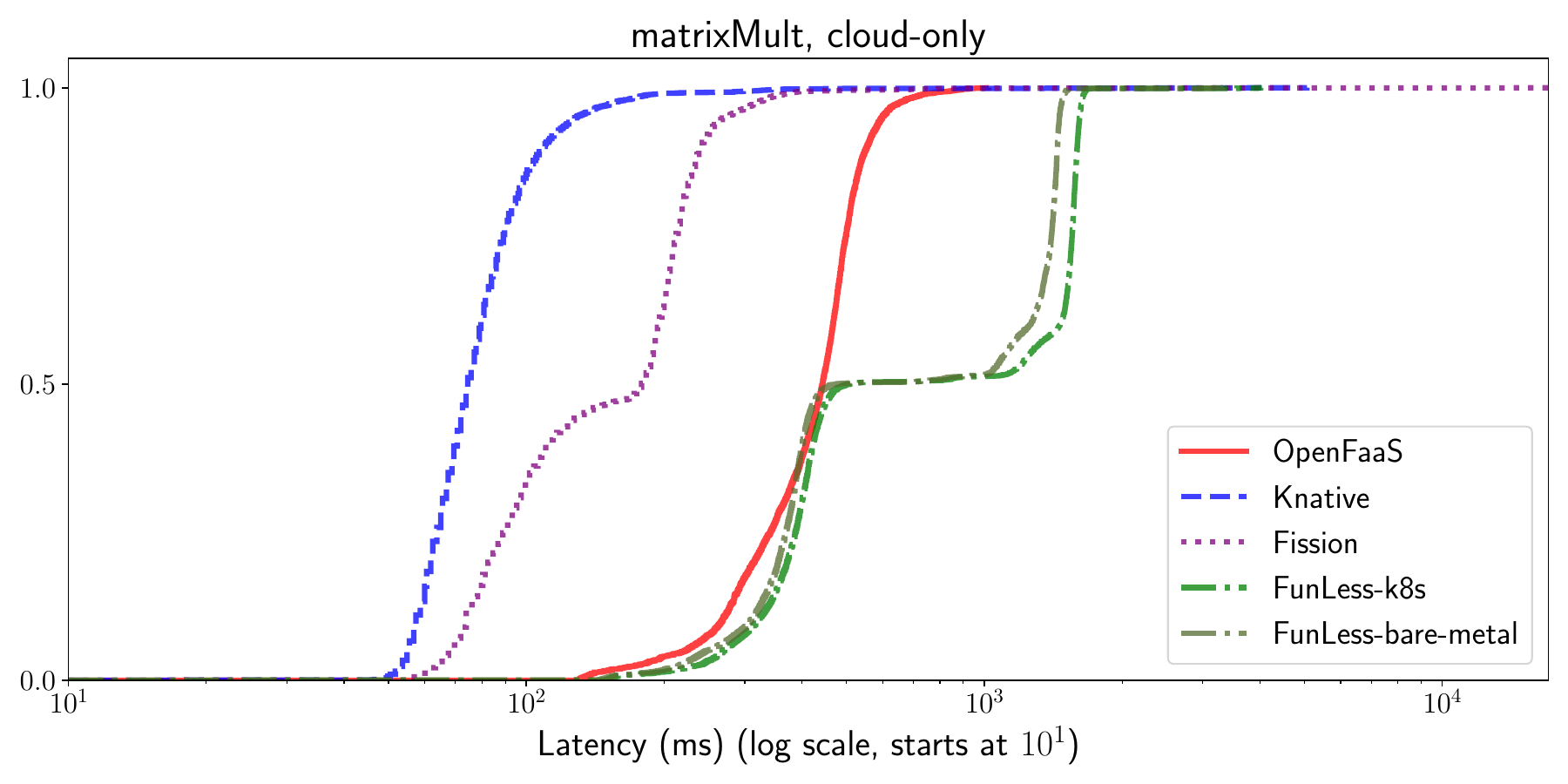}
    \includegraphics[width=.49\textwidth]{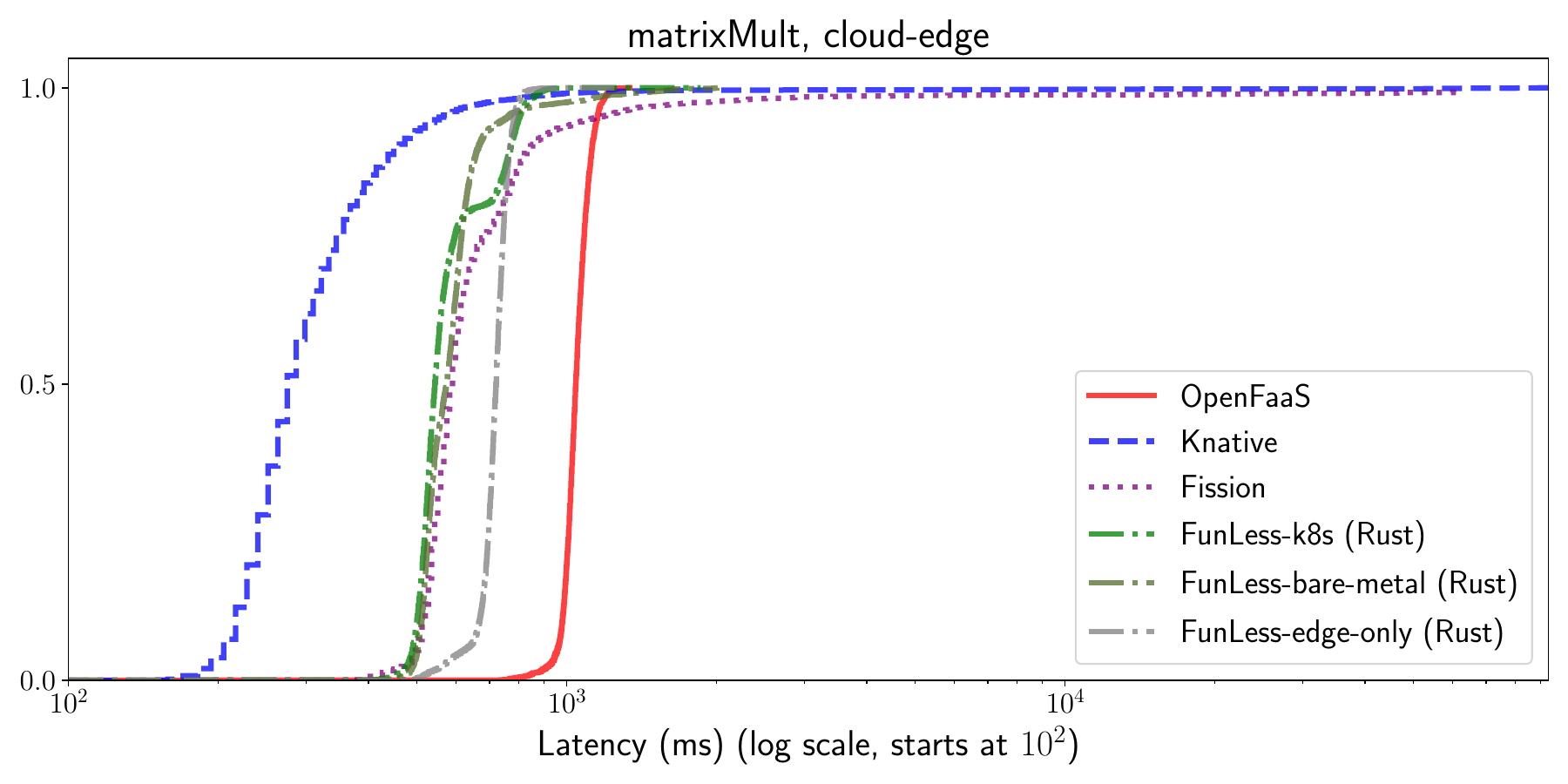}
    \includegraphics[width=.49\textwidth]{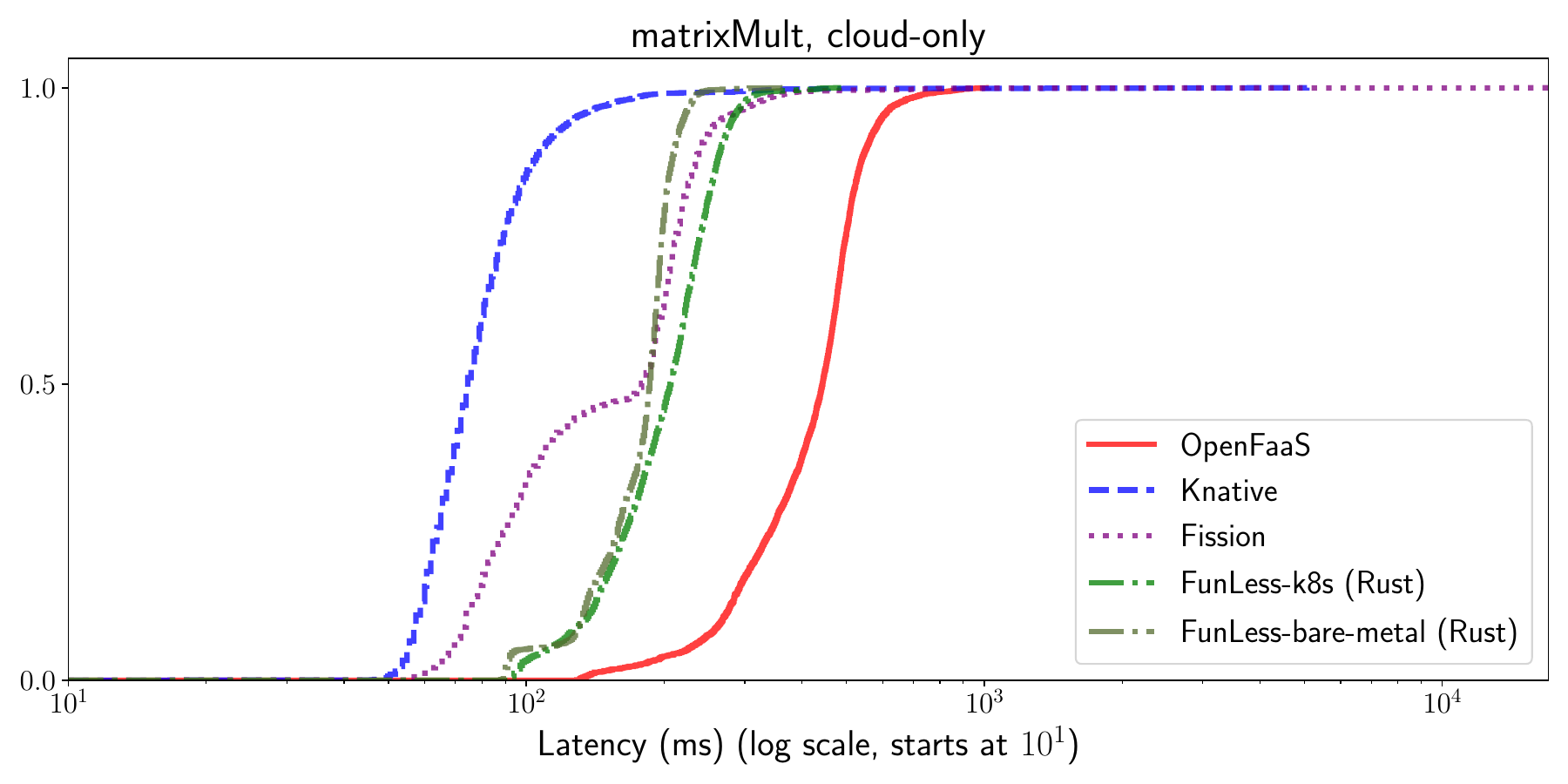}
    \caption{Top, the cumulative distribution of the latencies of the
        \textit{matrixMult} benchmark, \textit{cloud-edge} on the left and
        \textit{cloud-only} on the right. Bottom, the same plots except that FunLess
        runs an alternative version of the \textit{matrixMult} function written in a Rust instead of JS.}
    \label{fig:matrixmult}
\end{figure*}



We report the last benchmark, \textit{matrixmult}, in
\cref{fig:matrixmult}---for brevity, we report, also in this case, the
performance of \textsf{FunLess-edge-only} in the \textit{cloud-edge} plots. The
figure shows four plots instead of two because we run the same benchmark for
FunLess with an alternative implementation of the function in Rust to
investigate higher-than-expected latencies for FunLess. Specifically,
\cref{fig:matrixmult} reports at the top row the cumulative distribution of the
latencies of the \textit{matrixMult} benchmark---\emph{cloud-edge} on the left
and \emph{cloud-only on} the right---and at the bottom row the same distribution
except that FunLess runs an alternative version of the function written in a
Rust instead of JS. Indeed, looking at the top row of \cref{fig:matrixmult}, we
notice that FunLess is the worst-performing platform in all deployments and
scenarios. In particular, we note the ``step'' in FunLess' plots, which
corresponds to almost half of the requests having a higher latency. Since we
observe this phenomenon for both \textsf{Funless-k8s} and
\textsf{Funless-bare-edge}, we can discard the hypothesis that the issue comes
from the usage of Docker/Kubernetes.
To assess whether this problem derives from the JS/Wasm runtime or from some
peculiarity of the platform, we use an alternative implementation of the
\textit{matrixMult} function in Rust and plot the performance of this version at
the bottom of \cref{fig:matrixmult}. We stress that we do not use these
additional results to compare the performance of FunLess against the other
platforms for the \textit{matrixMult} case---which would be unfair because we do
not run the same Rust function on the other platforms since some do not support this language---but only
to investigate if the performance issue is linked with the usage of JavaScript.
%
Looking at the bottom of \cref{fig:matrixmult}, we see that FunLess obtains
%
smaller and more consistent latencies, which leads us to posit that
the issue lies indeed in the JavaScript runtime used by the current implementation of
FunLess.

Summarising the results from the four benchmarks, we conclude that FunLess
generally has performance comparable with the other platforms, especially when deployed
at the edge, without the support/weight of containerisation technologies.
Besides the results for the \textit{cloud-edge} and \textit{cloud-only}
configurations, we underline that only FunLess can run in an edge-only
configuration.

Before moving to the results on memory occupation, we underline that, while
reported in our results, one cannot directly compare the performance of FunLess
in the \textit{edge-only} and \textit{cloud-edge} configurations.
Indeed, in \textit{edge-only}, the \textit{Core} is in the same local network of
JMeter while, in the other configurations, the \textit{Core} is much farther,
hosted in the cloud. Moreover, in \textit{edge-only}, the \textit{Core} has much
fewer resources than when it runs in the cloud. We deem devising other
configurations that remove these differences and make for a fair comparison
between the different FunLess deployment modalities interesting future work.


\begin{figure}[h]
    \centering
    \includegraphics[width=0.9\linewidth]{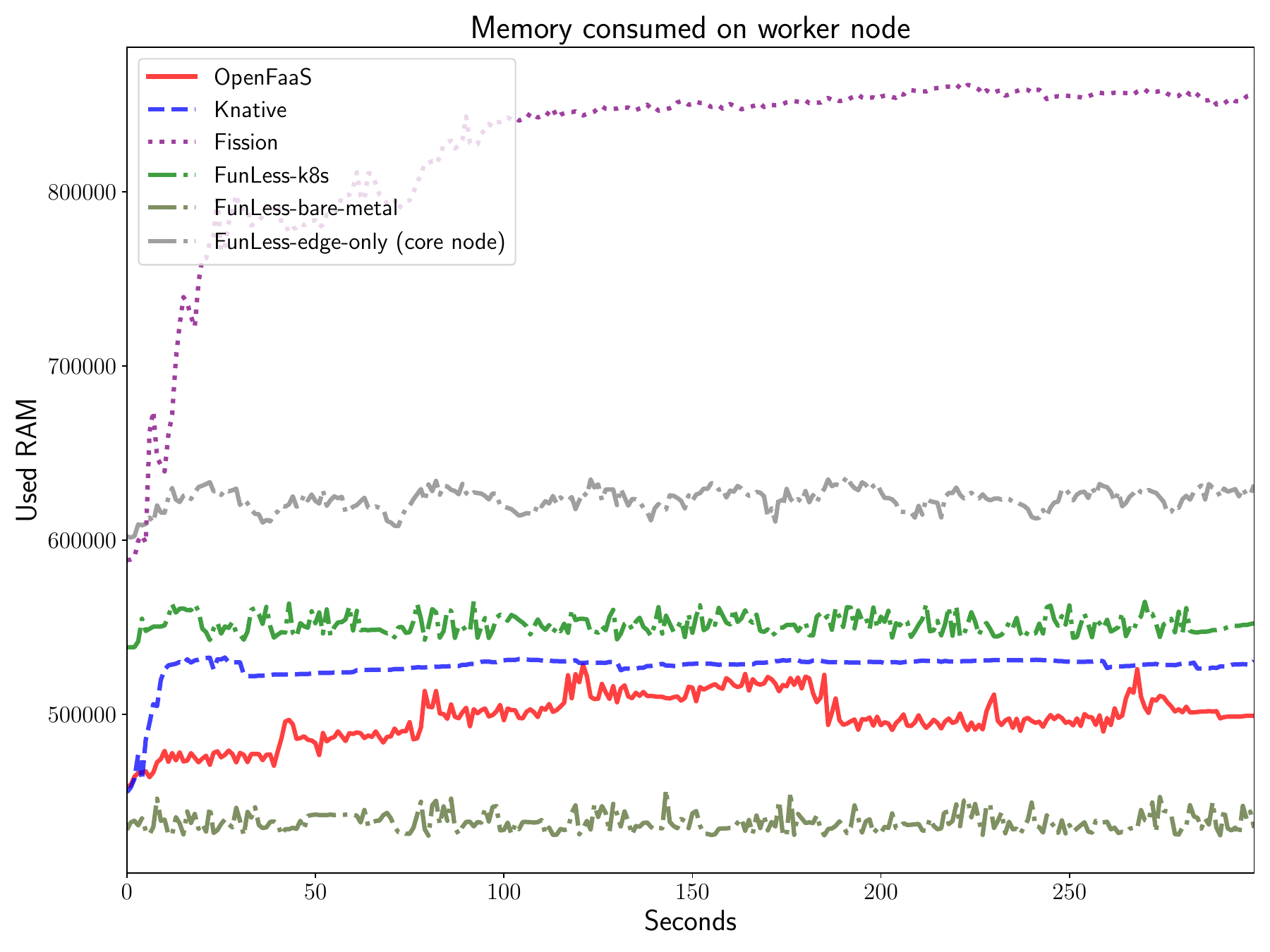}
    \caption{Memory consumption of the \textit{hellojs} function running on a Raspberry Pi 3B+.}
    \label{fig:mem_usage}
\end{figure}

We end this section by reporting in \cref{fig:mem_usage} the memory consumption
of the workers of the considered platforms (hence, not the core/controller) on
the edge devices, including the memory required by the operating system (ca.\@
300MB). For compactness and to provide a comprehensive view of the memory
consumption profile of FunLess, we plotted in \cref{fig:mem_usage} also the
consumption of FunLess' \textit{Core} in managing the invocations of the
\textit{hellojs} benchmark, labelling the plot line \textsf{FunLess-edge-only
(core node)}. When comparing the performance of FunLess against those of the
alternatives we omit to consider the results of this plot line because they
focus on just one component (the core/controller) of the platform. We comment on
this plot line at the end of this section.

From the results, Fission requires the highest amount of memory. This is a
consequence of the usage of the container pool used by the platform to reduce
cold starts. While \textsf{FunLess-k8s} is the second-highest in terms of memory
requirements (due to the stacking of the BEAM, Docker, and Kubernetes runtimes),
\textsf{FunLess-bare-edge}
requires the least amount of memory out of all the platforms (on average ca.\@
438 MB). Intuitively, this configuration can reach such a low memory footprint
because it omits the overhead generated by the services provided by Docker and
Kubernetes. In general, we note that, while FunLess allows one to deploy the
platform without the support of containers and container orchestrators, all the
considered alternatives heavily rely on the latter, making it unfeasible to
avoid their usage and prevent the overhead they generate. On the contrary, the
deployment flexibility afforded by FunLess allows one to have a functioning
\textit{Worker} running with minimal overhead (e.g., that of the underlying
operating system).

Additionally, looking at the plot trends, we notice that the amount of memory
required by FunLess, regardless of the deployment configuration, is relatively
stable compared to the alternatives. Technically, we explain this result because
the running \textit{Worker} requires a little more memory than the one it
allocated at launch (i.e., the memory at time 0, before the series of
invocations seen afterwards) plus the occupancy of the loaded Wasm modules in
the \textit{Worker}. Hence, the difference in the memory footprint of the
\textit{Worker} between its started/idle state and being under load is low and
mainly due to the memory consumption of the BEAM instance.

We contrast FunLess' predictable memory-consumption behaviour with OpenFaaS,
Knative, and Fission (the latter being the one where the phenomenon is mostly
accentuated), which can use considerably more memory between their start/idle
and under-load states, despite the simplicity and the minimal memory occupancy
of the \textit{hellojs} function. We quantify this trend also in terms of
standard deviation. The memory usages of \textsf{FunLess-k8s} and
\textsf{FunLess-bare-metal} has a standard deviation of 5,421 and 5,363 KB
respectively; slightly over half of that of Knative (9,502 KB), less than half
that of OpenFaaS (13,760 KB), and far lower than that of Fission (52,536 KB).

Finally, we look at the plot line of FunLess' \textit{Core} component, labelled
\textsf{FunLess-edge-only (core node)} in \cref{fig:mem_usage}. The
\textit{Core} uses around 620 MB, including the operating system, the database
(Postgres), the monitoring service (Prometheus) and Docker (used to deploy
Postgres and Prometheus), for an additional memory overhead of ca.\@ 450MB. We
chose to use Docker in this instance to simplify the deployment of the database
and the monitoring system. Of course, this configuration is not related to
FunLess and one could further reduce the overhead by deploying the whole stack
without containers. 

Overall, we deem these results on the memory consumption of FunLess appropriate
to have functions run on memory-constrained edge devices.

\section{Related Work}
\label{sec:related_work}

Looking at the work most closely related to our proposal, we find several
technological proposals targetting edge and cloud scenarios of serverless
systems, which Cassel et al.\@ thoroughly reviewed in a recent
survey~\cite{GVRMAC22}. Most of the solutions (86\%) for the IoT/edge field rely
on some container technology while promising technologies like WebAssembly and
Unikernels represent only 2-3\% of the proposals.


We are not aware of any other serverless framework that supports a Wasm runtime and allows both the controller and the workers to be deployed on resource-constrained edge nodes.

Focusing on serverless platforms supporting Wasm runtimes, Hall and Ramachandran were among the first to advocate WebAssembly as the enabling technology
to avoid the overhead of containers, which would substantially weigh on the
limited hardware resources of edge computing environments~\cite{HR19}. They presented a serverless platform that runs WebAssembly code
within the V8 JavaScript engine for execution and sandboxing of functions.
Differently from FunLess they use a
NodeJS runtime which embeds V8 for the execution of WebAssembly code. As the
authors note~\cite{HR19}, the nesting of these layers takes a conspicuous toll
on the performance of the system.


Gadepalli et al.~\cite{GMPCP20} use WebAssembly to run and sandbox
serverless functions. They target only single-host deployments, requiring the deployment of the entire platform on one node only.
Moreover, they do not support WASI~\cite{web:wasi}, thus making their system potentially less portable.

Gackstatter et al.~\cite{GFD22} propose WOW, a WebAssembly-based runtime
environment for serverless edge computing
integrated within the Apache OpenWhisk platform. The authors introduce a new
layer between OpenWhisk and different Wasm runtimes which enable the execution
of Wasm functions. Compared to FunLess, WOW requires the deployment of a full
installation (of a custom version) of the OpenWhisk platform which precludes the installation of the controller to low-power and memory-restricted edge devices.\footnote{We tried to deploy WOW on a multi-host cloud configuration for comparison purposes. Unfortunately, the deployment failed (the platform relies on an old and modified version of OpenWhisk that is not supported anymore, i.e., the last commit in the project is older than 2 years).}



Lucet~\cite{web:lucet} was used by Fastly to run Wasm on their commercial Compute platform.
%
Lucet translated WebAssembly to native code, which was then executed using Lucet's runtime also on edge devices. Unfortunately, Lucet has reached end-of-life and is no longer maintained.
Cloudflare Workers~\cite{CF_workers}
is also a commercial serverless platform that supports the possibility of defining functions in Wasm and has native support for WASI since 2022.\footnote{\url{https://blog.cloudflare.com/announcing-wasi-on-workers}}
Although the runtime part of this project has recently been made open-source,\footnote{\url{https://blog.cloudflare.com/workerd-open-source-workers-runtime/}},
the serverless platform is proprietary and closed-source.



It is worth mentioning the work by Shillaker and Pietzuch~\cite{SP20} that,
tangential to our proposal, concerns a Wasm-based serverless runtime that uses
Wasm to achieve state sharing across functions---they allow the execution of
functions that share memory regions in the same address space for possible
performance benefits. On a similar note, Zhao et al.~\cite{ZXCZZL23} present an
OpenWhisk extension for confidential serverless computing that integrates a
Wasm runtime. The authors propose a solution to construct reusable
enclaves that enable rapid enclave reset and robust security to reduce cold
start times. Although these kinds of proposals are orthogonal to FunLess, we see
them as interesting future optimisations that the usage of a Wasm function
runtime can unlock for FunLess.

Kjorveziroski and Filiposka~\cite{KF23b} focus on serverless orchestration using
Wasm and introduce a variant of Kubernetes that can orchestrate Wasm modules
that are executed without containers.
Interestingly, also Kjorveziroski and Filiposka report that Wasm tasks enjoy
faster deployment times (two-fold) and at least one order of magnitude smaller
artefact sizes, while still offering comparable execution performance.




Finally, Tzenetopoulos et al.~\cite{tzenetopoulos2021FaaSCuriousEdge} analyse
the performance of \textit{Lean OpenWhisk}, an edge-focused variant of the
Apache OpenWhisk serverless platform. Their variant of the platform coalesces
the scheduling and execution components in a single entity, remove the message
broker (Apache Kafka) from the deployment, and introduce changes to reduce
OpenWhisk's overhead, making it better suited for resource-constrained devices.
Unfortunately, we could not perform a quantitative comparison of Lean OpenWhisk
because the project seems no longer maintained, and we could not deploy the
platform due to problems linked to years-old container images (NodeJS 6) and outdated
dependencies that reached the end of life (Ansible 2.7.9).

\section{Conclusion}
\label{sec:conclusion}
We present FunLess, a FaaS platform tailored to respond to recent trends in
serverless computing that advocate for extending FaaS to cover private edge
cloud systems, including Internet-of-Things devices. The motivation behind the
shift towards private edge cloud systems includes reduced latency, enhanced
security, and improved resource usage.
Unlike existing solutions that rely on containers and container orchestration
technologies for function invocation, FunLess leverages Wasm as its
function-execution runtime environment. The reason behind this choice is to
reduce performance overheads that can prevent resource-constrained devices from
running FaaS systems. Wasm's fundamental feature exploited by FunLess is its
lightweight, sandboxed runtime, which allows the platform to run efficiently
functions in isolation on constrained devices at the edge. Thus, Wasm provides a
portable, homogeneous way for developers to implement and deploy their functions
among clusters of heterogeneous devices (write once, run everywhere),
simplifying platform deployments, offering flexibility in deployment options,
and mitigating cold start issues.

To validate FunLess, we consider different deployment scenarios spanning from
the pure edge case to the pure cloud one, with/without container orchestration
technologies to ease the deployment. To draw our comparison, we select three
alternatives from production-ready, widely adopted open-source FaaS
platforms---OpenFaaS, Fission, and Knative---and run representative cloud and
edge FaaS benchmarks.

Our benchmarks confirm that FunLess is a viable solution for FaaS private edge
cloud systems. Our platform outperforms the considered alternatives in terms of
memory footprint and support for heterogeneous devices.


As future work,
we plan to integrate the new versions of Wasmtime and, with it, native support
for HTTP and other optimisations and features of the new releases and support
for the WASI runtime. Indeed, many current Wasm runtime implementations miss
features like interface types, networking support in WASI
multi-threading, atomics, and garbage collectors. Besides Wasmtime, other
projects are developing new, optimised, and extended Wasm runtimes, which
FunLess can leverage to increase its performance (and adapt it to different
application contexts). As an example, we conjecture that by using a Wasm runtime that natively supports HTTP
requests, FunLess will perform better in the network benchmark (c.f.\@
\cref{evaluation}) while the support for garbage collection can support improved JavaScript runtimes and increase the performance of the matrix
multiplication benchmark.




We also plan to improve the reliability of the platform, allowing the support of
retry policies for failed invocations, at-least-once message delivery, and the
replication of \textit{Core} components. Following the principles of simplicity
and versatility that guided the development of FunLess, we propose to tackle
these extensions as optional features to support flexible deployments, adaptable to the different application contexts (cloud, edge, on resource-constrained devices).

Finally, we plan to ease the deployment of FunLess 
by supporting other deployment tools like, e.g., Nomad~\cite{web:nomad} 
and optimise the platform for edge devices by using, e.g., 
Nerves\cite{web:nerves} to further minimise the overhead on our bare-metal deployment.


\bibliographystyle{IEEEtran} 
\bibliography{biblio}

\end{document}